\documentclass[9pt,twocolumn,twoside]{osajnl}
\journal{ao} 
\usepackage{amsmath,amssymb,graphicx,paralist,epstopdf,subfigure}

\setboolean{shortarticle}{false} 

\title{Smear correction of highly-variable, frame-transfer-CCD images with application to polarimetry}

\author[1,*]{Francisco A. Iglesias}
\author[1]{Alex Feller}
\author[1,2]{K. Nagaraju}

\affil[1]{Max Planck Institute for Solar System Research, Justus-von-Liebig-Weg 3, 37077  G\"{o}ttingen, Germany }
\affil[2]{Indian Instiute of Astrophysics, Koramangala Second Block, Bengaluru-560034, India }

\affil[*]{Corresponding author: iglesias@mps.mpg.de}

\dates{Compiled \today}

\ociscodes{(040.1520) CCD, charge-coupled device; (100.3020) Image reconstruction-restoration; (110.5405) Polarimetric imaging}

\doi{\url{http://dx.doi.org/10.1364/ao.XX.XXXXXX}}

\begin{abstract}Image smear, produced by the shutter-less operation of frame transfer CCD detectors, can be detrimental for many imaging applications. Existing algorithms used to numerically remove smear, do not contemplate cases where intensity levels change considerably between consecutive frame exposures. In this report we reformulate the smearing model to include specific variations of the sensor illumination. The corresponding desmearing expression and its noise properties are also presented and demonstrated in the context of fast imaging polarimetry. 
\end{abstract}

\setboolean{displaycopyright}{true}

\begin{document}

\maketitle
\thispagestyle{fancy}
\ifthenelse{\boolean{shortarticle}}{\abscontent}{}

\section{Introduction}
\label{sec:introduction}

CCD sensors with frame transfer architecture are currently of wide spread use in many imaging applications. A \textit{frame transfer CCD} collects photo-charges in a light sensitive area during exposure time to then rapidly clock them towards a shielded, frame storage area where readout is performed \cite{ccd}. This kind of detector readily become an interesting candidate when frame rate, duty cycle and optical fill factor need to be simultaneously maximized.
 
If the CCD is continuously  illuminated during frame transfer, the resulting image will be smeared reducing its contrast and spatial resolution \cite{ccd}. This artifact can be a draw back in many applications \cite{dorrington2005, choi2004, legrand2012, zhao2014} becoming particularly relevant when high frame rate and duty cycle are required. Given its general applicability, low cost and null effect on duty cycle among others, numeric algorithms for post-facto correction of image smear are an attractive alternative to the classic mechanical or optoelectronic shutter systems. 

The first desmearing algorithms were introduced by Powell \textit{et al.} \cite{powell1999}, for standard, charge flush and reverse clocking CCD operation modes, and Ruyten \cite{ruyten1999} only for charge flush mode. In both of the above-mentioned works authors assumed, among others, that pixel saturation has not taken place, noise in the images is negligible and the image illuminating the CCD does not change within exposure and frame transfer time. The first of these assumptions was investigated by Knox \cite{knox2007}, who showed that smearing can be used in specific cases to recover the values of saturated pixels in the image. More recently, different smearing correction techniques have been developed that are optimal for specific imaging applications \cite{sun2012, gao2011} or used for real time correction \cite{han2009}.

In all the above-mentioned works, the original assumption in \cite{powell1999, knox2007} of constant illumination is maintained. In this report we firstly develop the smearing model to treat cases where sensor illumination cannot be considered constant between consecutive frames, provided the photon flux is constant between transitions, and the transitions occur in synchronization with the frame transfers. Further we assume the flux transition profile is symmetric with respect to its inflection point (Section \ref{sec:model}). Secondly we study the corresponding desmearing algorithm and its conditioning for the cases of non-periodic and periodic variable scenes (Section \ref{sec:rest}).

The results of this work may be of utility  to any application where the scene is highly changing in synchronization with the detector readout. An example are the fast polarimetric  measurements  we use to demonstrate our results (Section \ref{sec:application}).

\section{Frame transfer model with variable illumination}
\label{sec:model}
We will assume frame transfer is performed in columns direction and neither pixel saturation nor blooming has occurred. Under these assumptions the analyses in the present and following sections are restricted to a single sensor column without losing generality.

\begin{figure}[h!]
\centerline{\includegraphics[width=0.9\columnwidth]{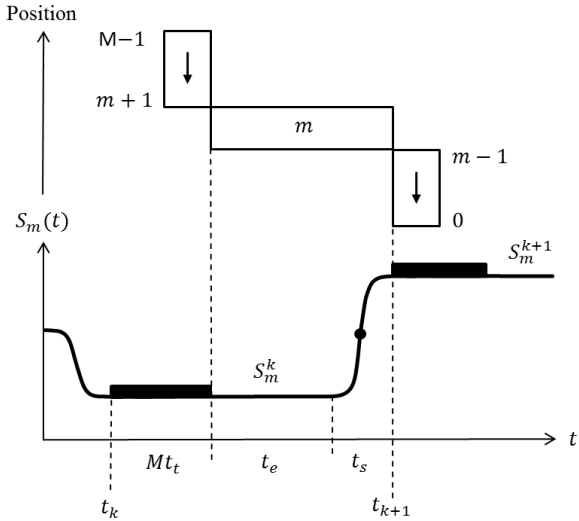}}
\caption{Process required to acquire pixel no. \textit{m} of frame no. \textit{k}, in a frame transfer CCD with \textit{M} rows operating in standard mode. Upper sketch illustrates the position within the column of charge well no. \textit{m}  during frame acquisition. The lower profile exemplifies the model we adopted for the photo charges flux at pixel no. \textit{m}, $S_m(t)$. Namely constant during frame transfer (black boxes with duration $Mt_t$) and exposure time ($t_e$), and variable during switching time ($t_s$). The transition profile is assumed symmetric with respect to the point (black circle) with coordinates $\{t_{k+1}-t_s/2,  (S_m^k+S_m^{k+1})/2\}$. See the text for additional details.}
\label{fig:fig1}
\end{figure}

 Fig. \ref{fig:fig1} illustrates the operation in \textit{standard mode} of a CCD with \textit{M} rows during the acquisition of a single pixel in row no. \textit{m} of frame no. \textit{k}. The upper sketch represents the position of charge well no. \textit{m} within the column. The lower profile exemplifies the values of the \textit{photo charges flux} in pixel no. \textit{m} as function of time, $S_m(t)$. The considered frame extends from time instants $t_k$ to $t_{k+1}$ with $t_{k+1}-t_k$ being the reciprocal of the CCD frame rate. The sensor is assumed to be permanently illuminated. The black boxes delimit the intervals where frame transfers are performed.

Acquisition of frame no. $k$ starts during the transfer of frame no. $k-1$. An empty charge well is shifted from the top of the array to position no. \textit{m} during a fraction of the total frame transfer time given by $(M-m+1) t_t$, where $t_t$ is the \textit{period of the charge transfer clock}. After this, static accumulation at position no. \textit{m} takes place during the \textit{exposure time}, $t_e$. The flux is assumed to be constant and equal to $S_m^k$ during frame transfer and exposure time.  A transition of the flux to its constant value in the next frame, $S_m^{k+1}$, is then contemplated. The flux transition occurs during the \textit{switching time} $t_s$, with the charge well still at position no. \textit{m}. The flux transition profile is assumed to be symmetric with respect to the point with Cartesian coordinates $\{t_{k+1}-t_s/2, (S_m^k+S_m^{k+1})/2\}$ in Fig. \ref{fig:fig1} (black circle). In the final step, once the flux is stable again, the charge well is transfered to the storage area of the sensor. 

For the above-described process and neglecting noise contributions (see Sections \ref{sec:model}\ref{subsec:noisemodel} and \ref{sec:rest}\ref{subsec:restnoper}), pixel saturation, charge transfer inefficiencies and instabilities of the charge transfer clock, the value of the smeared signal acquired in pixel no. \textit{m} of frame no. \textit{k} can be expressed as,
\begin{multline}
\hat{Y}_m^k={{r_1 t_t}} \sum_{j=m+1}^{M-1}g_j S_j^k+g_m t_e S_m^k + {1\over2}g_m t_s (S_m^k+S_m^{k+1})+\\
d_m(t_e+t_s) +b_m+{r_2 t_t} \sum_{j=0}^{m-1}g_j S_j^{k+1}+{t_t} \sum_{\substack{j=0 \\ j\neq m}}^{M-1}d_j,
\label{eq:eq1}
\end{multline} 
where the time constants $g_j$, $d_j$ and $b_j$ denote the  \textit{gain},  \textit{dark current flux} and  \textit{bias} of pixel no. \textit{j} respectively. The two ad-hoc coefficients $r_1$ and $r_2$, have been introduced to facilitate experimental tuning of the model, e.g. to: contemplate possible differences in the photo-charges-generation efficiencies of the static and transferring clocking states; compensate the effects of synchronization errors between the CCD readout and flux switching; model other CCD operation modes; etc.

The simple form of the third term in the right hand side of Eq. \ref{eq:eq1}, which accounts for the signal accumulated during $t_s$, is a direct consequence of the  symmetry restriction imposed on the flux transition profile. The later has been defined by the application that motivated this work. In spite of that, notice it also covers other cases of possible practical interest like very fast switching (step-like profile), linear switching (ramp-like profile), constant illumination, etc.

In the following we will assume the acquired images have been offset corrected using the dark frame,
\begin{equation}
\hat{D}_m=d_m(t_e+t_s) +b_m+{t_t} \sum_{\substack{j=0 \\ j\neq m}}^{M-1}d_j,
\label{eq:eq3}
\end{equation}
which in practice can be obtained by averaging a long sequence of dark images, i.e. with $S_m(t)=0$.

In addition, let us define the \textit{unsmeared signal} acquired during the static and constant illumination exposure as,
\begin{equation}
Y_m^k=g_m t_e S_m^k
\label{eq:eq2}
\end{equation}
and constants,
\begin{align}
\alpha= {t_s\over{2 t_e}}, && \delta_1= {r_1 t_t\over{t_e}} && \text{and} && \delta_2 = {r_2 t_t\over{t_e}}.
\label{eq:eq2b}
\end{align}
Note that Eq. \ref{eq:eq2} implies any gain table corrections should be applied after the desmearing process (see Section \ref{sec:rest}).

Using Eq. \ref{eq:eq1}-\ref{eq:eq2b} and redefining $\hat{Y}_m^k-\hat{D}_m$ as $\hat{Y}_m^k $, we can write a smeared column of the sensor  in matrix form as,
\begin{equation}
\hat{\bold Y}^ k= \bold{A Y}^k + \bold{B Y}^{k+1},
\label{eq:eq5}
\end{equation}
where, $\mathbf{\hat{Y}}^k=\begin{bmatrix} \hat{Y}_0^k , \hat{Y}_1^k ,\dots, \hat{Y}_{M-1}^k \end{bmatrix}^T, \mathbf Y^k=\begin{bmatrix} Y_0^k,Y_1^k ,\dots , Y_{M-1}^k \end{bmatrix}^T,$
\begin{multline*}
\mathbf A= 
\begin{bmatrix}
1+\alpha &  \delta_1 & \dots &  \delta_1   \\
0 & 1+\alpha & \dots &  \delta_1  \\
 \vdots & \vdots & \ddots  & \vdots \\
0 & 0& 0 & 1+\alpha \\
\end{bmatrix}, 
\mathbf B= 
\begin{bmatrix}
\alpha &  0 & \dots & 0   \\
\delta_2 & \alpha & \dots &  0  \\
 \vdots & \vdots & \ddots  & \vdots \\
\delta_2 & \delta_2 & \delta_2 & \alpha \\
\end{bmatrix},
\end{multline*}
and the superscript $T$ denotes the transpose.

 Note that, an analogue analysis can be carried out for the case where frame transfer takes place \textit{before} the flux switching, deriving in a \textit{causal} difference equation also with matrix coefficients  $\bold A$ and  $\bold B$. 
 
In case the value of $t_t$  it is not known and can not be directly measured, some techniques have been developed to allow its experimental estimation for specific scenarios \cite{ruyten1999,feng2012}.

When operating the CCD in \textit{charge flush mode}, the charge wells of the light sensitive area are flushed after the frame transfer is finished. Eq. \ref{eq:eq5} can be modified to model this scenario by doing $r_1=0$. In \textit{reverse clocking mode} the wells of the light sensitive area are transfered to a charge drain, located at the top of the sensor, by means of a reverse clocking. The latter is done after the usual frame transfer. To consider this case in Eq. \ref{eq:eq5}, $\bold A$ has to be replaced by $\bold A^T$ and $r_1$ and $r_2$ adjusted appropriately if the clock rates, or charge transfer inefficiencies, of the usual frame transfer and reverse sweep are different.

\subsection{Noise properties of the smeared images}
\label{subsec:noisemodel}
To investigate the noise properties of the smearing process, we consider the unsmeared columns in Eq. \ref{eq:eq5} as independent random vectors. We also include an independent, additive noise term, $\mathbf{N}^k$ to the right hand side of Eq. \ref{eq:eq5} to account for extra noise sources added during or after the charge wells are read, e.g. readout noise. For the above-described scenario one can show that,
\begin{equation}
  \begin{split}
\mathrm{Var}[\mathbf{\hat Y}^k]=\mathbf A\mathrm{Var}[\mathbf{Y}^k]\mathbf{A}^T+\mathbf B\mathrm{Var}[\mathbf{Y}^{k+1}]\mathbf{B}^T+\mathrm{Var}[\mathbf{N}^k]\\
\mathrm{Cov}[\mathbf{\hat Y}^k,\mathbf{\hat Y}^{k+1}]=\mathbf B\mathrm{Var}[\mathbf{Y}^{k+1}]\mathbf{A}^T,
\label{eq:nmodel}
  \end{split}
\end{equation}
where $\mathrm{Var}$ and $\mathrm{Cov}$ represent the variance-covariance and cross-covariance operations respectively. Note that Eq. \ref{eq:nmodel} evidences the spatial and temporal correlation introduced by the smearing process. 

We further assume in Eq. \ref{eq:nmodel}, that $\mathrm{Var}[\mathbf{Y^{k+1}}]=\gamma \mathrm{Var}[\mathbf{Y^{k}}]$ with $\gamma \in \mathbb{R}$, e.g. due to photon noise variation between frames with very different illumination levels. Then, a simple upper limit for the noise degradation can be written as follows,
\begin{equation}
\mid\mathrm{Var}[\mathbf{\hat Y^k}]\mid \leq \eta \mid\mathrm{Var}[\mathbf{Y^k}]\mid+\mid\mathrm{Var}[\mathbf{N^k}]\mid,
\label{eq:ulnmodel}
\end{equation}
where $\mid .\mid$ denotes the euclidean matrix norm and,
\begin{equation}
\eta=(1+\alpha+2M\delta_1/\pi)^2+\gamma(\alpha+2M\delta_2/\pi)^2 \text{ for } M\gg1.
\label{eq:chi}
\end{equation}
In the latter we used the fact that $\mathbf A=(1+\alpha-\delta_1)\mathbf I+\delta_1 \mathbf\Delta$ and $\mathbf B=(\alpha-\delta_2)\mathbf I+\delta_2 \mathbf \Delta^T$, where $\mathbf I$ is the $MxM$ identity matrix and $\mathbf \Delta$ an $MxM$ upper triangular matrix with $\Delta_{i,j}=1$ for $j\geq i$ and $\Delta_{i,j}=0$ for $j<i$.

\begin{figure}
\centerline{\includegraphics[width=1\columnwidth]{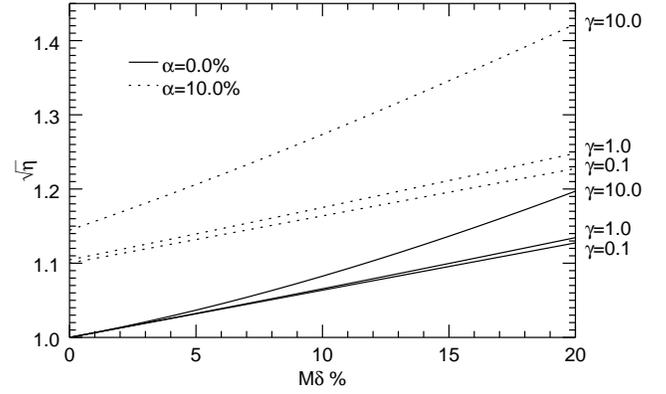}}
\caption{Upper boundary for the noise degradation due to the smearing process, versus the parameters product $M\delta$. The curves in each set, continuous for $\alpha=0.0\%$ and dotted for $\alpha=10.0\%$, correspond to $\gamma =0.1, 1.0$ and $10.0$ from bottom to top respectively. See Eq. \protect{\ref{eq:ulnmodel}-\ref{eq:chi}} for extra details.}
\label{fig:ulnmodel}
\end{figure}

Fig. \ref{fig:ulnmodel} shows curves of $\sqrt\eta$ versus $M\delta$ for different values of $\alpha$ and $\gamma$ (with $r_1=r_2=1$). When interpreting Figure \ref{fig:ulnmodel}, and similar ones in the next section, recall that by definition for a fixed frame rate and exposure time, the sum $2\alpha+M\delta$ is constant. 

\section{Image restoration}
\label{sec:rest}
\subsection{Non-periodic illumination}
\label{subsec:restnoper}
In this section we discuss the solution of Eq. \ref{eq:eq5} to an arbitrary time series of smeared columns, $\mathbf{\hat Y}=\begin{bmatrix}(\mathbf{\hat Y}^0)^T,(\mathbf{\hat Y}^1)^T,\dots,(\mathbf{\hat Y}^{K-1})^T \end{bmatrix}^T$, with $K\geq 2 \in \mathbb{N}$. Given the shape of matrices $\bold A$ and $\bold B$, the restored columns, $\mathbf{\bar Y}=\begin{bmatrix}(\mathbf{\bar Y}^0)^T,(\mathbf{\bar Y}^1)^T,\dots,(\mathbf{\bar Y}^{K-1})^T \end{bmatrix}^T$, can be efficiently obtained by solving iteratively from pixel no. $M-1$ to $0$ and frame no. $K-1$ to $0$ if the final condition, $\mathbf{\bar Y}^{K}$, is known. 

\begin{figure}
\centerline{\includegraphics[width=1\columnwidth]{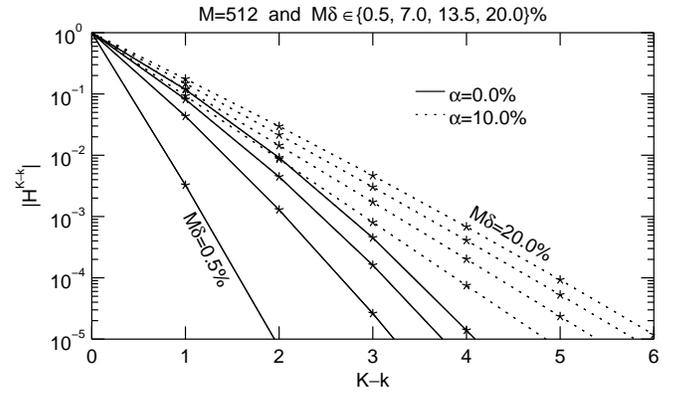}}
\caption{Upper boundary for the propagation of errors in the final condition of  Eq. \protect{\ref{eq:eq5}}, versus index of the restored frame. The curves in each set, continuous for $\alpha=0.0\%$ and dotted for $\alpha=10.0\%$, correspond to $M\delta =0.5, 7.0, 13.5$ and $20.0\%$ from left to right respectively}. Note the vertical axis is in logarithmic scale. See  Eq. \protect{\ref{eq:gsol}} for extra details.
\label{fig:hsol}
\end{figure}

The difficulties to obtain $\mathbf{\bar Y}^{K}$ can be avoided after considering the general solution of Eq. \ref{eq:eq5}, 
\begin{equation}
\mathbf{\bar Y}^k=\sum_{j=1}^{K-k}\mathbf H^{K-k-j}\mathbf A^{-1}\mathbf{\hat Y}^{K-j} + \mathbf H^{K-k}\mathbf{\bar Y}^K,
\label{eq:gsol}
\end{equation}
where $\mathbf H^k=(-\mathbf A^{-1}\mathbf B)^k$, has a highly damped  homogeneous term for practical values of its parameters. Note that we use upper scripts of time independent matrices to denote exponentiation. Fig. \ref{fig:hsol} presents $\mid \mathbf H^{K-k}\mid$ as function of $K-k$. The curves were computed for $M=512$, although they are practically insensitive to $M$ for $M>>1$, and different values of $\alpha$ and $\delta$ (with $r_1=r_2=1$). As can be appreciated,  the error introduced by a wrong estimation of the final condition, e.g. $\mathbf{\bar Y}^{K}=\mathbf{\hat Y}^{K-1}$, can be reduced to negligible levels provided few frames at the end of the series are dropped.

To study the overall conditioning of the restoration algorithm, with respect to uncertainties in the measured columns, let us write it using block matrices as follows (see Eq. \ref{eq:gsol}),
\begin{equation}
\mathbf{\bar Y}=\mathbf U \mathbf{\hat Y}+\mathbf H\mathbf{\bar Y}^K,
\label{eq:gsolbm}
\end{equation}
where $\mathbf H=\begin{bmatrix}(\mathbf H^K)^T, (\mathbf H^{K-1})^T,\dots,(\mathbf H^1)^T\end{bmatrix}^T$,
\begin{equation*}
\mathbf U= 
\begin{bmatrix}
\mathbf A^{-1} & \mathbf H^1 \mathbf A^{-1}  & \mathbf H^2 \mathbf A^{-1} &  \dots & \mathbf H^{K-1}\mathbf A^{-1}   \\
\mathbf 0 & \mathbf A^{-1}  &  \mathbf H^1 \mathbf A^{-1} &  \dots & \mathbf H^{K-2}\mathbf A^{-1}   \\
 \vdots & \vdots  & \vdots & \ddots  & \vdots \\
\mathbf 0 & \mathbf 0 & \mathbf 0  & \dots  & \mathbf A^{-1} \\
\end{bmatrix},
\end{equation*}
and $\mathbf 0$ is the $M\times M$ null matrix.

\begin{figure}
\centerline{\includegraphics[width=1\columnwidth]{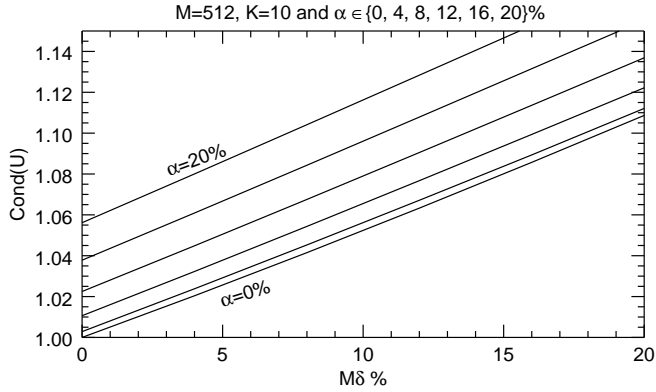}}
\caption{Conditioning of the desmearing algorithm with respect to errors in the measured images, versus the parameters product $M\delta$. The curves correspond to $\alpha =0, 4, 8, 12, 16$ and $20\%$ from bottom to top respectively. See Eq. \protect{\ref{eq:gsolbm}} for extra details.}
\label{fig:errprop}
\end{figure}

According to Eq. \ref{eq:gsolbm}, when neglecting the contributions of $\mathbf H\mathbf{\bar Y}^K$ (see Fig. \ref{fig:hsol}), the propagation of relative errors in the measured $ \mathbf{\hat Y}$ is upper bounded by the condition number of $\mathbf U$, i.e. the ratio of the largest to the smallest singular values of $\mathbf U$ \cite{num_recipes2007}. The latter is plotted in Fig. \ref{fig:errprop} versus $M\delta$ (with $r_1=r_2=1$)  for $M=512$, $K=10$ and different values of $\alpha$.  The curves are practically insensitive to $M$ and $K$ for $M>>1$ and $K>>1$. As expected, the restoration becomes monotonically worse conditioned for increasing $\delta$ and  $\alpha$.

\subsection{Periodic illumination}
\label{subsec:restper}
Of particular interest for this work is the solution of Eq. \ref{eq:eq5} to a time-periodic series of smeared columns, i.e. the illumination pattern repeats cyclically after a fixed amount of frames. Assuming periodicity in the scene allows the usage of image accumulation for noise reduction purposes. This is specially important for high-frame-rate applications that are photon starved. 

In contrast to Section \ref{sec:rest}\ref{subsec:restnoper}, in the periodic illumination scenario,  image restoration can be performed for a single period of the input,  $\mathbf{\hat Y}=\begin{bmatrix}(\mathbf{\hat Y}^0)^T,(\mathbf{\hat Y}^1)^T,\dots,(\mathbf{\hat Y}^{\tilde K-1})^T \end{bmatrix}^T$, where $\tilde K\geq 2 \in \mathbb{N}$ is the period. The corresponding desmearing expression can be written in block matrix as, 
\begin{equation}
\mathbf{\bar Y}=\mathbf{\tilde U}^{-1} \mathbf{\hat Y},
\label{eq:persol}
\end{equation}
where,
\begin{equation*}
\mathbf{\tilde U}= 
\begin{bmatrix}
\mathbf A &  \mathbf B & \mathbf 0 & \dots & \mathbf 0   \\
\mathbf 0 & \mathbf A &  \mathbf B & \dots & \mathbf 0   \\
 \vdots & \vdots  & \vdots & \ddots & \vdots \\
\mathbf B & \mathbf 0 & \mathbf 0 &\dots &\mathbf A  \\
\end{bmatrix},
\end{equation*}
is a $M\tilde K\times M\tilde K$, block circulant matrix \cite{mazancourt1983}. Note Eq. \ref{eq:persol} is linear in $\mathbf{\hat Y}$ and thus it can be applied after averaging any number of periods, reducing this way the computational load.

On one hand, numeric inversion of $\mathbf{\tilde U}$ will not be in general a problem because it has to be done only once for each setup, i.e. values of $\delta_1$, $\delta_2$ and $\alpha$. There are, however, Fourier-based methods used to invert block circulant matrices that can considerably reduce computation time and storage while increasing numeric stability \cite{mazancourt1983,vescovo1997}. 

On the other hand, the product in the right side of  Eq. \ref{eq:persol} may become computationally expensive because it has to be performed once per image column. Two possible approaches to cope with this issue are: find only an approximate solution iteratively by first guessing $\mathbf{\hat Y}^0$ similar to Section  \ref{sec:rest}\ref{subsec:restnoper}; or use the fact that $\mathbf{\tilde U}^{-1}$ is also block circulant and band dominant \cite{vescovo1997} to implement an efficient multiplication algorithm.

\section{Application to fast imaging polarimetry}
\label{sec:application}
We applied the desmearing expression given in Section  \ref{sec:rest}\ref{subsec:restper} to correct a set of images taken by the Fast Solar Polarimeter (FSP) \cite{feller2014}. We will not describe the instrument details here, only the relevant aspects related to the image smearing process. 

FSP uses a 264x264 pixels, split-frame-transfer pnCCD \cite{hartman2006} that can record up to 800 frames per second (fps). The pnCCD operates in standard mode and its readout is synchronized with a four-states beam modulator \cite{keller2003} that will introduce an intensity variation according to the polarization state of the incoming light. The intensity switching profiles and synchronization between pnCCD and modulator meet the assumptions detailed in section \ref{sec:model} for Eq. \ref{eq:eq5}.

Any complete polarimetric measurement requires taking at least four consecutive exposures, one during each modulator state, that can have very different light levels for strongly polarized sources. The four recorded frames are linearly combined in a later step to retrieve the polarization signal. The sensor is constantly illuminated during measurements. If the incoming polarization state is constant, a periodic series of four images is obtained.

\begin{figure}
\centerline{\includegraphics[width=1\columnwidth]{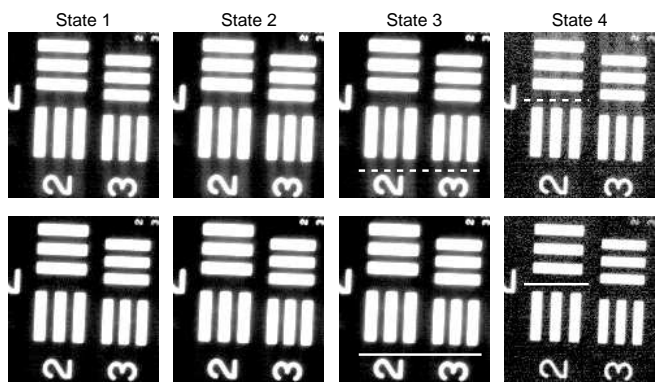}}
\caption{Smeared (upper row) and corrected (lower row) images, displayed using logarithmic scales to enhance the smearing. The gray scales of the images in each column are the same. The mean values of the bright areas in the upper-row images are 1950, 2828, 2825 and 297 arbitrary counts for states 1 to 4 respectively. The dashed and continuous horizontal lines in states 3 and 4 images,  delimit the cuts shown in Fig. \protect\ref{fig:results2}}.
\label{fig:results}
\end{figure}

The upper row of Fig. \ref{fig:results} presents an example measurement taken at 700 fps of a USAF target illuminated by a light source of constant linear polarization. Only a fraction of the total sensor area is shown. Each of the four exposures, denoted with labels State 1 to State 4, corresponds to an average of several measurement periods. Smearing is obvious in all images, note in particular its variable spatial distribution among different states. The latter can be easily explained using Eq. \ref{eq:eq5} and considering that the mean light levels in the bright areas, i.e. above two sigmas from the image mean, of each state are very different. Namely 1950, 2828, 2825 and 297 arbitrary counts approximately for state 1 to state 4 respectively. 

The lower row of Fig. \ref{fig:results} displays the restored images obtained after applying Eq. \ref{eq:persol} with $\tilde K=4$, $M=264$, $\alpha=3.9\%$, $\delta_1=0.05\%$ and $\delta_2=0.03\%$ to the images in the upper row. Reduction of the smearing is evident with only faint residuals. 

\begin{figure}
   \centering 
   \subfigure{
      \includegraphics[width=1\columnwidth]{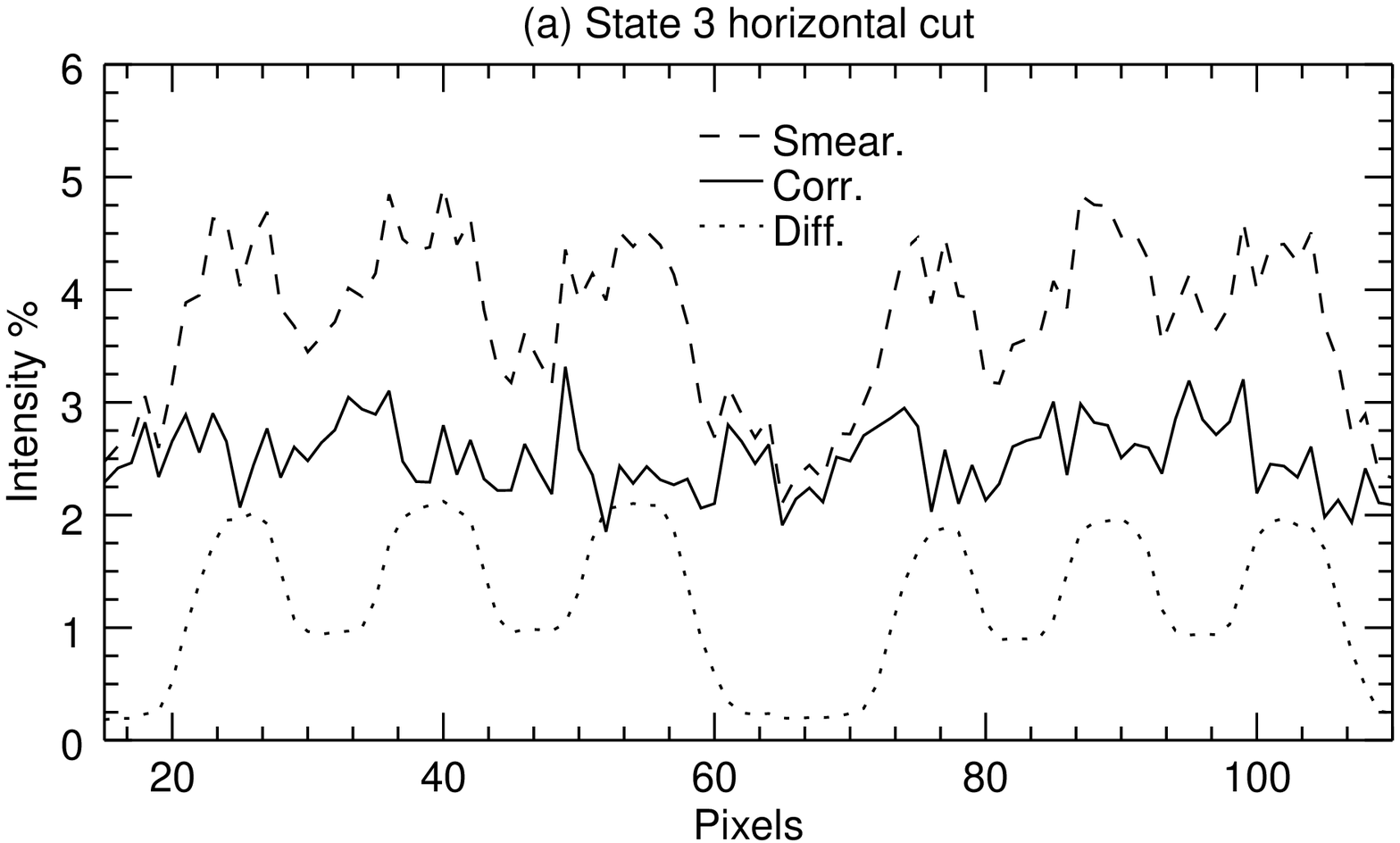}
      \label{fig:results2a}
    }
   \subfigure{
      \includegraphics[width=1\columnwidth]{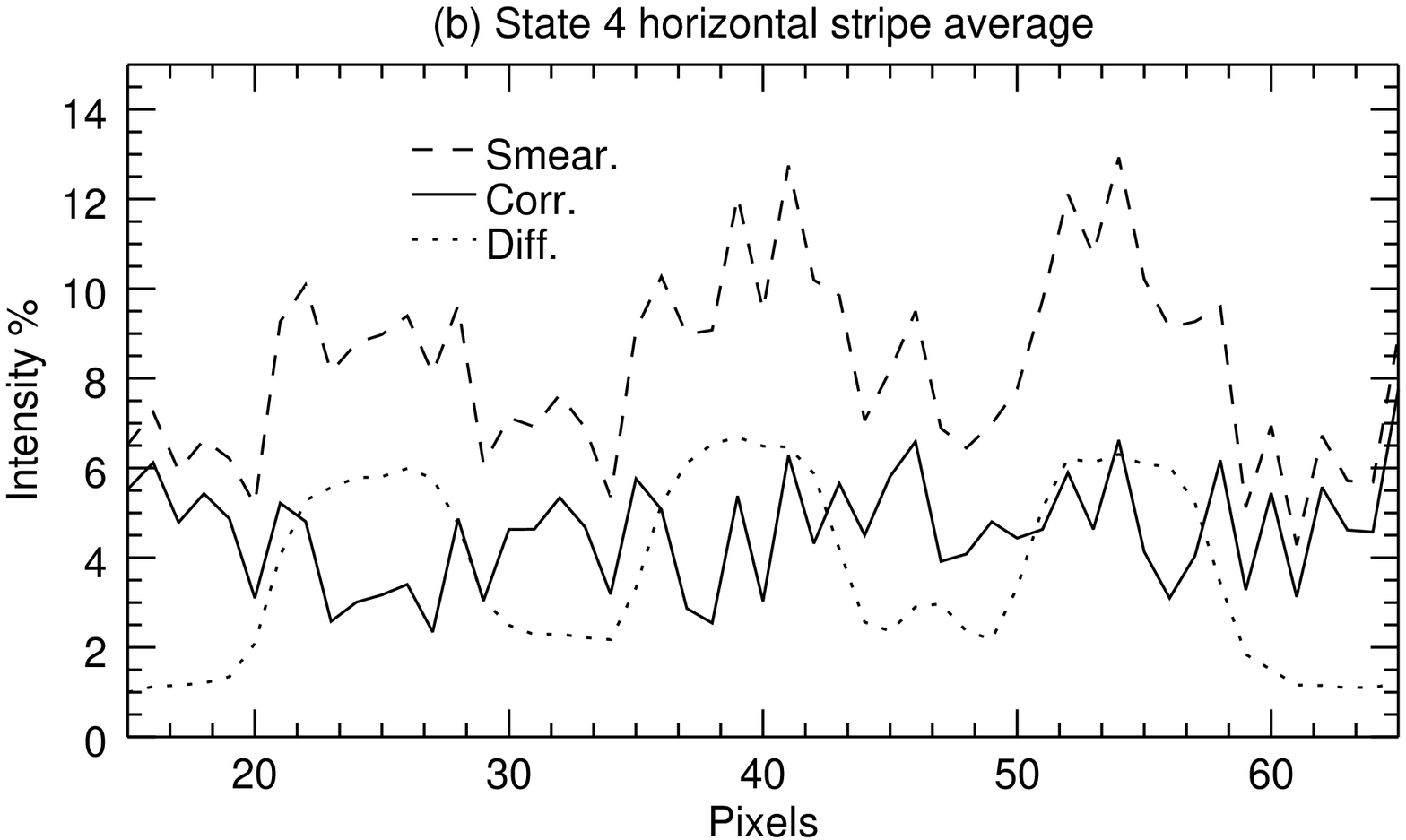}
      \label{fig:results2b}
    }    
\caption{Horizontal cut of the state 3 images (a) and vertical average of a 4-pixel-wide horizontal stripe of state 4 images (b). The position of the cut and the center of stripe are indicated by the dashed and continuous, white lines in Fig. \protect\ref{fig:results}. In both plots the values for the smeared (dashed) and corrected (continuous) images, as well as their difference (dotted) are shown. All profiles are expressed in percentage of the mean value in the bright areas of the corresponding smeared image.}
\label{fig:results2}
\end{figure}

The quality of the restoration and the differences in the artifact levels between state 3 and state 4 images, are exemplified by the profiles shown in Fig. \ref{fig:results2}. Firstly, from the corrected profiles, note there is not clear residual smearing above the noise levels. Given the USAF target pattern, and in complete absence of smearing and stray light, the profiles of the selected cuts are expected to be flat except for noise induced variations.

Secondly, from the smeared profiles in Fig. \ref{fig:results2}, note that the maximum artifact levels, relative to the mean in the bright areas of each image, are larger in State 4 ($\sim 12\%$) than in State 3 ($\sim 5\%$). To understand this, recall that each frame accumulates smearing signal during \textit{both} the pre-exposure and the post-exposure frame transfers (see black boxes in Fig. \ref{fig:fig1}). Further, note that the smearing signal acquired during the post-exposure transfer depends on the illumination level of the following frame. In this way, smearing level in State 3 is low because the consecutive frame (state 4) has approximately 9.5 times fainter illumination. On the contrary, the artifact level in State 4 is high because the consecutive frame (State 1, recall  the input is periodic) has approximately 6.6 times larger intensity.

\section{Conclusions}
\label{sec:conclusions}
We developed a model for the smearing, introduced in frame transfer CCDs, that accounts for variations in the sensor illumination provided the changes take place in synchronization with the detector readout and have a transition profile that is an odd function. The derived model depends only on three physical parameters, namely the period of the charge transfer clock, the image exposure time and the photon flux switching time. 

In addition, we showed that smearing introduces not only a spatial but also a temporal correlation in the images and derived a simple upper boundary for the noise degradation.

We also studied the corresponding desmearing algorithm and its conditioning for the cases of non-periodic and periodic variable scenes. The latter was successfully applied to restore a set of polarimetric measurements taken by the Fast Solar Polarimeter. 

The present work does not take in to account the effects of pixel saturation, pixel blooming, and charge transfer inefficiencies.

F. A. Iglesias would like to acknowledge the International Max Planck Research School for Solar System Science for supporting his participation in this project. 

\bibliographystyle{osajnl}
\bibliography{Iglesias_et_al_2015}

\end{document}